\documentclass[aps,pra,epsfigure,superscriptaddress,twocolumn,longbibliography]{revtex4-1}
\usepackage{dcolumn}
\usepackage{bm}
\usepackage{graphicx}
\usepackage{amsmath}
\usepackage{latexsym}
\usepackage{amsfonts}
\usepackage{amssymb}
\usepackage{array}
\usepackage{epsfig}
\usepackage{txfonts}
\usepackage{color}
\usepackage[colorlinks=true,linkcolor=blue,urlcolor=blue,citecolor=blue,pdfusetitle]{hyperref}
\usepackage{bm}
\usepackage{setspace}
\usepackage{braket}
\usepackage[utf8]{inputenc}
\usepackage[percent]{overpic}
\usepackage{xcolor}

\newcommand{\beq}{\begin{equation}}
\newcommand{\eeq}{\end{equation}}
\newcommand{\beqa}{\begin{eqnarray}}
\newcommand{\eeqa}{\end{eqnarray}}

\newcommand{\mossy}[1]{{\color{black}#1}}

\begin{document}

\title{An efficient non-linear Feshbach engine}

\author{Jing Li}
\affiliation{Department of Physics, Shanghai University, 200444, Shanghai, People's Republic of China}

\author{Thom\'as Fogarty}
\affiliation{Quantum Systems Unit, Okinawa Institute of Science and Technology Graduate University, Okinawa, 904-0495, Japan}

\author{Steve Campbell}
\affiliation{Instituto Nazionale di Fisica Nucleare, Sezione di Milano, \& Dipartimento di Fisica, Universit{\`a} degli Studi di Milano, Via Celoria 16, 20133 Milan, Italy}

\author{Xi Chen}
\email{xchen@shu.edu.cn}
\affiliation{Department of Physics, Shanghai University, 200444, Shanghai, People's Republic of China}

\author{Thomas Busch}
\email{thomas.busch@oist.jp}
\affiliation{Quantum Systems Unit, Okinawa Institute of Science and Technology Graduate University, Okinawa, 904-0495, Japan}
\date{\today }

\begin{abstract}
We investigate a thermodynamic cycle using a Bose-Einstein condensate with nonlinear interactions as the working medium. Exploiting Feshbach resonances to change the interaction strength of the BEC allows us to produce work by expanding and compressing the gas. To ensure a large power output from this engine these strokes must be performed on a short timescale, however such non-adiabatic strokes can create irreversible work which degrades the engine's efficiency. To combat this, we design a shortcut to adiabaticity which can achieve an adiabatic-like evolution within a finite time, therefore significantly reducing the out-of-equilibrium excitations in the BEC. We investigate the effect of the shortcut to adiabaticity on the efficiency and power output of the engine and show that the tunable nonlinearity strength, modulated by Feshbach resonances, serves as a useful tool to enhance the system's performance.
\end{abstract}
\maketitle

\section{Introduction}
\label{intro}

The ability to accurately control quantum systems using the latest available experimental techniques has opened the door to thinking about experiments which can study the nature of thermodynamics in the quantum realm~\cite{CampisiRMP,JohnReview,FuscoPRX,Thermo1}. In particular, it is interesting to study conceptual heat engines, which are systems governed by a complex dynamics and in which the manipulation of quantum states can be used to produce work. Due to the recent progress in precisely controlling the non-adiabatic dynamics of ultracold atoms and Bose-Einstein condensates (BECs) \cite{BEC5,BEC1,BEC2,BEC6,BEC7},
and the ability to tune many of their system parameters, they offer ideal systems with which to create quantum heat engines \cite{Hallwood,LutzPRLHeatEngine,LutzScience}.

To do work in a heat cycle, the Hamiltonian must be modified as adiabatically as possible to avoid unwanted excitations. This, however, requires long timescales which can have drawbacks, such as exceeding the life time of the of cold atom systems and having a low output power. One method to circumvent this issue is to use a shortcut to adiabaticity (STA) \cite{MugaJPB,ChenPRL104}, which allows for fast quantum state manipulation while also suppressing final excitations 
 \cite{SchaffPRA,SchaffEPL,SchaffNJP}. Techniques such as quantum transitionless driving (or counter-diabatic driving), fast-foward algorithms, and inverse engineering have been shown to yield states with high fidelity in finite time, see the recent reviews~ \cite{STAreview,PRXAdolf}. While these approaches have often centered
around non-interacting systems, STAs have also been explored in interacting, nonlinear, and other systems~\cite{adolfo1,interacting,CampbellPRL,David,Stringari,JingSciRep}. Remarkably, STAs
have fundamental implications on quantum speed limits \cite{QSL_MT,QSL_ML,QSLDeffner,QSLreview,DeffnerReview}, time-energy uncertainty relations (or energy cost) \cite{Chenenergy,costXi,CampbellPRA,SantosSciRep,CampbellPRL2,Ecut,SantosFICT,SarandyArXiv}, and the quantification of the third law of thermodynamics in the context of
quantum refrigerators \cite{EPL09,EPL11}, which results in intriguing practical applications in heat engines~\cite{GooldSciRep,GongPRE,delCampoEntropy,LutzArXiv,LutzArXiv2,FunoPRL,Kosloff2017}.

\begin{figure}[t]
\includegraphics[width=0.9\columnwidth]{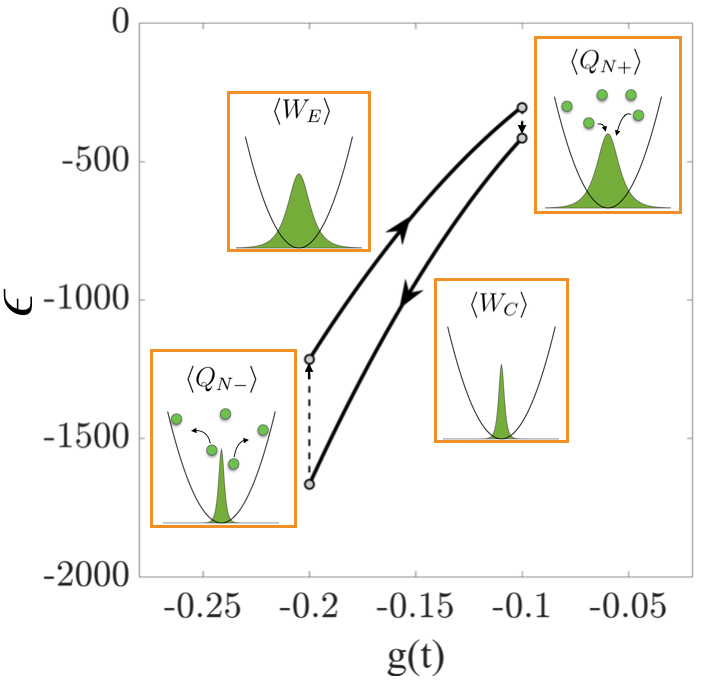}
\caption{Otto cycle: Work is done during the compression $\langle W_C \rangle$ and expansion strokes $\langle W_E \rangle$. Through coupling to an external reservoir the number of particles, $N$, can be increased or decreased, thereby dissipating particles $\langle Q_{N-}\rangle$ and absorbing particles $\langle Q_{N+}\rangle$ and therefore increasing or decreasing the energy.}
\label{schematic}
\end{figure}

In this work we analyze a quantum Otto cycle \cite{ottoX,ottoQ,Lutz1,Lutz2,optotto,ottoCY,otto1,Brendan}, (see Fig. \ref{schematic}) that uses a BEC with attractive nonlinear interactions, which can be described as a bright soliton, as its working medium. The creation and control of bright solitons in BECs is well established experimentally~\cite{solitons,solitons2,solitons3,cornish2,cornish3,hulet}, making it an ideal nonlinear interacting system to investigate. Rather than using the STA to modulate the trapping potential and therefore do work on the system, as has been investigated previously for example in Refs.~\cite{GooldSciRep,delCampoEntropy,LutzArXiv,LutzArXiv2,GongPRE}, we instead fix the trapping frequency and compress and expand the soliton by time dependently varying the nonlinear interaction strength. This can be achieved experimentally by exploiting the Feshbach resonances of the BEC, \mossy{thereby realizing a form of Feshbach engine. Feshbach resonances are a powerful tool used in cold atom experiments, which allow to tune the scattering length of elastic collisions between atoms by using magnetic or optical fields \cite{Feshbach}. Such resonances, the thermodynamics of which were recently studied~\cite{MossyNJP}, occur when the energy of a bound state of an interatomic potential is equal to the energy of a pair of colliding atoms, resulting in the enhancement of interparticle interactions about the resonace point. To control the time dependence of the interaction, while ensuring suppressed excitations, we} use a recently designed STA for this system~\cite{JingSciRep} and investigate its effectiveness by comparing the performance with a suitably rescaled adiabatic modulation. We show that the STA approach leads to higher final target state fidelities and lower irreversible work. We further analyse the engines performance by calculating the efficiency and output power with respect to the cycle duration and find that while arbitrarily fast modulations are ineffective, the STA significantly enhances the overall performance on intermediate timescales. Finally, we highlight the remarkable role that the non-linear interaction strength plays, showing that due to the effect it has on the energy spectrum, stronger nonlinear interactions allows for increased performance.

The remainder of the paper is organized as follows. In Sec.~\ref{model} we present the model and define the thermodynamic quantities used throughout. Sec.~\ref{shortcut} briefly reviews the techniques used to design a STA for dynamically changing the scattering length for soliton matter waves (as devised in Ref.~\cite{JingSciRep}) and in Sec.~\ref{resultsSTA} we examine the performance of the STA during a compression.  The efficiency and power output of the quantum Otto cycle using the STA is analysed in Sec.~\ref{resultsENG} and in Sec.~\ref{conclusions} we present our conclusions.

\section{Model and Figures of Merit}
\label{model}
We consider the one-dimensional Gross-Pitaevskii equation describing the dynamics of a harmonically trapped BEC
\begin{equation}
\label{eq1}
\left[-\frac{1}{2}\psi_{xx}+g(t)|\psi(x,t)|^2+\frac{1}{2}x^2\right]\psi(x,t)=\mu(t) \psi(x,t),
\end{equation}
where $\psi(x,t)$ is the wave-function, $\mu(t)$ is the chemical potential, and $g(t)$ describes the non-linear interatomic interaction strength, which can be experimentally tuned by applying a Feshbach resonance. Here, we have adopted harmonic oscillator units such that all lengths are scaled by $\sqrt{\hbar/m\omega}$, time $t$ by $1/\omega$, and the interaction strength $g$ by $\sqrt{\hbar^3\omega/m}$, where $m$ is the mass of the condensate and $\omega$ is the frequency of the harmonic trapping potential.

In this work we will focus on attractive interactions, $g(t)<0$, where an exact solution of Eq.~(\ref{eq1}) in the absence of the harmonic trap is given by the well-known hyperbolic secant ansatz for a bright soliton matter-wave
\begin{equation}
\psi(x,t)=A(t)\mathrm{sech}\left(\frac{x}{a(t)}\right)\;.
\label{ansatz}
\end{equation}
Here, $A(t)$ is the amplitude and $a(t)$ is the width of the soliton, and the system is normalized with respect to the number of particles in the soliton $N=\!2aA^{2}=\!\int^{+\infty}_{-\infty}|\psi(x,t)|^{2} dx$ with $A\!=\!\sqrt{N/(2a)}$. As the width $a(t)$ depends on the interaction strength, varying $g(t)$ leads to compressions and expansions of the soliton. Even though Eq.~\eqref{ansatz} is the free-space solution, we will in the following assume that it is still a good approximation in weak trapping potentials.

Varying the interaction strength necessarily implies work being performed on/by the soliton \mossy{through a change in its energy}. The energy of the soliton is given by
\beq
\epsilon(t)=\int dx\left[ \frac{1}{2}\vert\nabla \psi(x,t)\vert^2 +\frac{1}{2}x^2\vert\psi(x,t)\vert^2 -\frac{g(t)}{2}\vert\psi(x,t)\vert^4 \right]\;,
\eeq
where we denote $\epsilon_{i(f)}$ as its initial (final) energy. It \mossy{therefore} follows that the work done during the process is given by the change in the energy
\beq
W(t)=\epsilon(t)-\epsilon_i.
\eeq
Furthermore, for a quasi-static process the average work done, $\langle W \rangle$, is equal to the \mossy{adiabatic energy change, $\langle W^{AD} \rangle$, which in the case of the system being at zero temperature is simply given by the} difference between the initial and final energies
\beq
\langle W \rangle = \mossy{\langle W^{AD} \rangle} = \epsilon_f-\epsilon_i \nonumber.
\eeq
Nonadiabatic processes require $\langle W \rangle\!\geq\!\mossy{\langle W^{AD} \rangle}  $, thus implying that a degree of irreversibility has been introduced. This can be quantified through the irreversible work
\beq
\label{eq:Wirr}
\langle W_{irr} \rangle=\langle W \rangle-\mossy{\langle W^{AD} \rangle}\;,
\eeq
\mossy{which in our case is created when applying the control pulses used to manipulate the soliton, as they transiently excite the system. We will use the above} quantities to assess the performance of a shortcut to adiabaticity applied to the manipulation of a soliton matter-wave, and its potential use as a small scale engine. For this engine we consider the well-studied Otto cycle as illustrated in Fig.~\ref{schematic},
however since our description of the soliton relies on solving the Gross-Pitaevskii equation which only describes the zero-temperature-state of the system, we make an analogy for the role of temperature in our system. In a physical setting, such a condensed soliton would be surrounded by thermal atoms which would add to the condensate fraction if subsequently cooled. Similarly the condensate fraction of the soliton would decrease with increasing temperature, thus removing particles from the soliton. \mossy{Situations where quantum bright solitons coexist with free thermal atoms in one-dimensional gases of attractive bosons have recently been a topic of large interest~\cite{castin,weiss1,weiss2,weiss3}}. We therefore envisage a particle engine where the compression and expansion strokes are controlled by the interaction strength $g(t)$ and do the work $\langle W_C \rangle$ and $\langle W_E \rangle$ respectively, and the final two strokes of the cycle are at fixed interaction while being coupled to external reservoirs which inserts energy $\langle Q_{N-} \rangle$ or extracts energy $\langle Q_{N+} \rangle$ by removing or adding particles to the soliton.

\section{Shortcuts to Adiabaticity for Soliton Matter Waves}
\label{shortcut}
Compressing or expanding the soliton in a short finite time can create irreversibility in the form out-of-equilibrium excitations, which will hamper the efficiency of the engine cycle. Therefore we aim to employ a STA which will suppress these excitations and ensure the final state has a large overlap with the one that would have been created in a fully adiabatic process. Such a technique was recently developed in Ref.~\cite{JingSciRep} and we briefly review it in this section.

For bright solitons the evolution of the width of the cloud, $a(t)$, is related to the non-linear interaction strength through
\begin{equation}
\label{aDiff}
\ddot{a}(t)+a(t)=\frac{4}{a^3(t) \pi^2}+\frac{2g(t)N}{\pi^2a^2(t)},
\end{equation}
which comes from the variational principle with the hyperbolic secant ansatz, Eq.~\eqref{ansatz}. To obtain the adiabatic limit, we can solve
$\ddot{a}=0$ from Eq. (\ref{aDiff}), to find the dependence of the soliton width on the interaction strength
\begin{equation}
\label{adiabticlimit}
a^4(t)-\frac{2g(t)N}{\pi^2}a(t)=\frac{4}{\pi^2},
\end{equation}
which gives an adiabatic reference for the soliton width in terms of the nonlinear interaction in the approximation of a weak trapping potential
\begin{equation}
\label{ADreference}
a_c(t)\simeq-\frac{2}{Ng_c(t)}\;.
\end{equation}
The above differential equation has a close analogy with the
dynamical equation of motion of a fictitious classic particle
with position $x$ in a perturbed Kepler problem
\begin{equation}
\label{kepler}
U(t)\simeq \frac{2g(t)N}{\pi^2 a(t)} +\frac{2}{\pi^2 a^2(t)}\;.
\end{equation}
It is worthwhile to note that $a_c (t)$ is indistinguishable from $a(t)$ found from Eq. (\ref{kepler}), corresponding to the minimal energy of Kepler potential, $\partial U(t)/\partial a(t) =0$.

Using a polynomial ansatz for $a(t)$ given as \mossy{$a_p(t)=\sum_{i=0}^5 a_i t^i$}, we can fix the  boundary conditions for the start, $t\!=\!0$, and end, $t\!=\!T_f$, of the stroke as \cite{JingSciRep}
\begin{equation}
\begin{aligned}
\label{boundarycondition}
& a_p(0) = a_c(0), & a_p(T_f) &= a_c(T_f), \\
&\dot{a}_p(0) = \dot{a}_c(0)=0,~~~ & \dot{a}_p(T_f) & = \dot{a}_c(T_f)=0, \\
&\ddot{a}_p(0) = \ddot{a}_c(0)=0,~~~& \ddot{a}_p(T_f) & = \ddot{a}_c(T_f)=0,
\end{aligned}
\end{equation}
by choosing a smooth adiabatic reference
$g_c(t)=(g_i+g_f)/2+9(g_i-g_f)\cos(\pi t/T_f)/16+(g_f-g_i)\cos(3\pi t/T_f)/16$.
Solving this set of equations allows us to determine the coefficients of the polynomial ansatz $a_p(t)$. Setting $a(t) \to a_p(t)$ in Eq.~\eqref{aDiff} and rearranging for $g(t)$ we arrive at the desired ramp of the non-linear interaction strength of a bright soliton matter-wave that realizes a STA for a chosen $T_f$.

\begin{figure}[t]
\hskip0.1\columnwidth(a)\hskip0.45\columnwidth {\bf (b)}
\includegraphics[width=0.48\columnwidth]{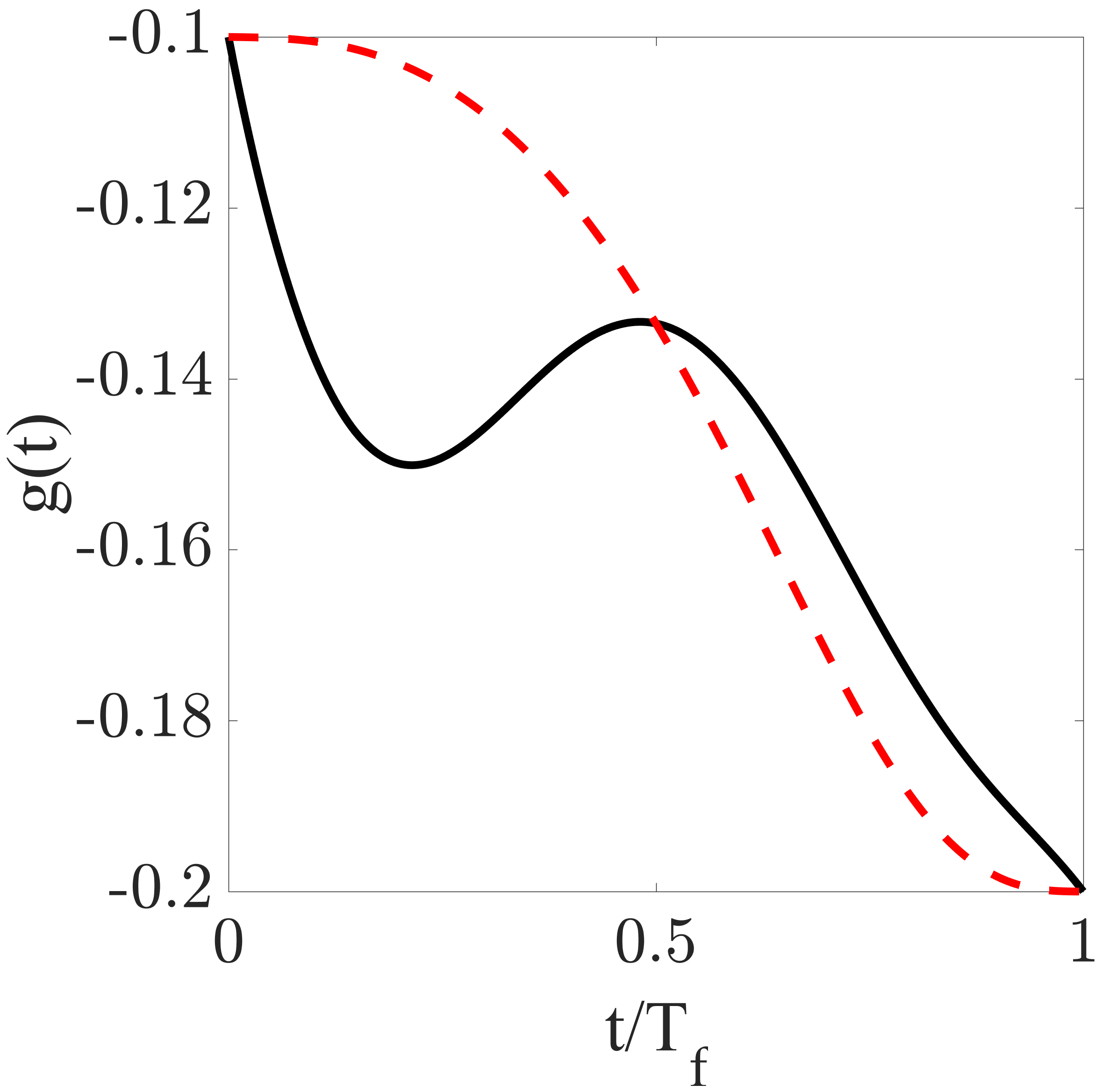}
\includegraphics[width=0.49\columnwidth]{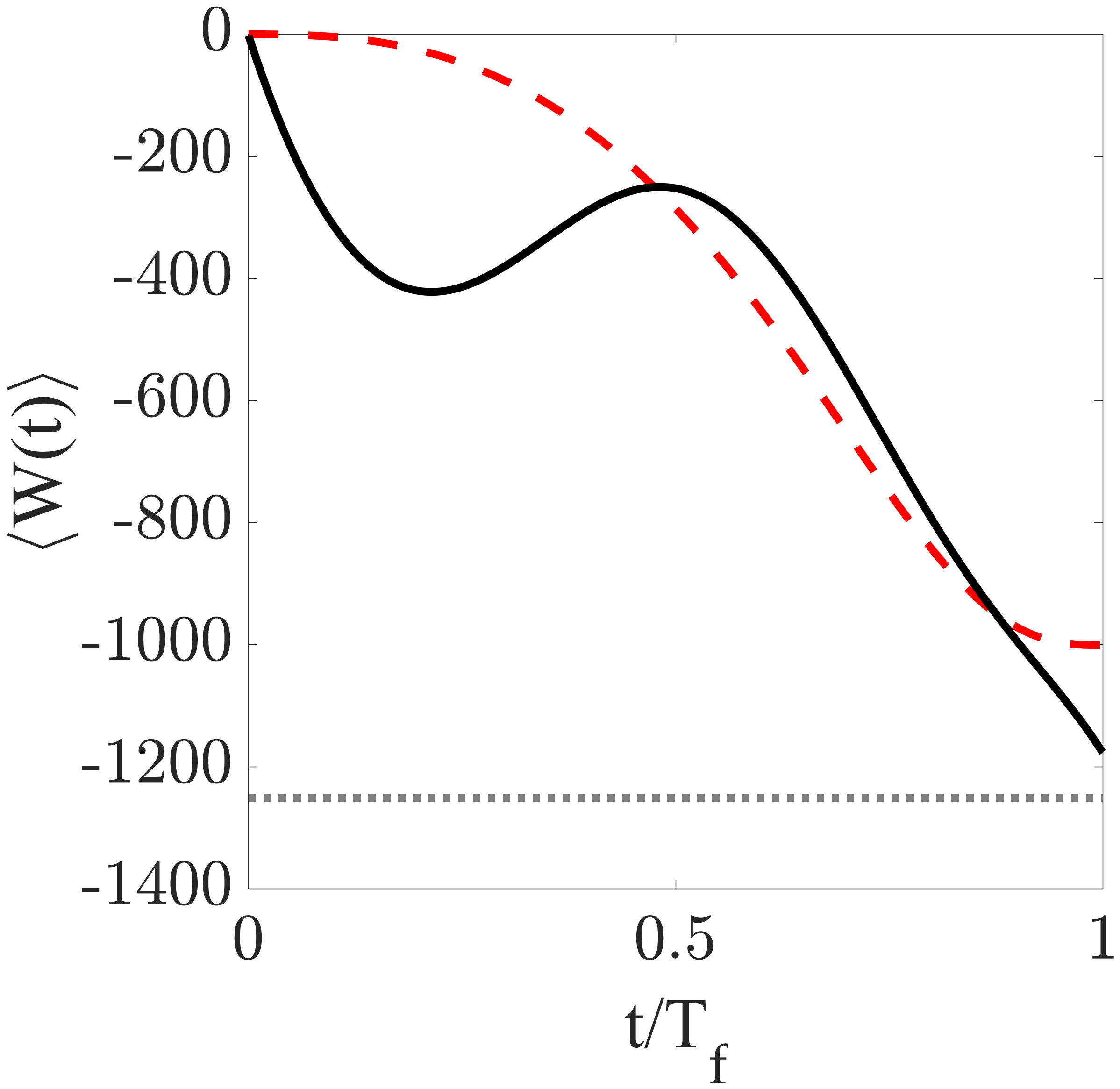}
\caption{(a) Modulation of the nonlinearity $g(t)$ for a compression stroke. We show the profile as designed by the STA for a finite time stroke $T_f=0.15$ (solid, black) and for a stroke in the adiabatic limit $ T_{AD}$ (dashed, red). We fix $N=100$. (b) Work done during the compression stroke ($t\leq T_f$), when the interaction strength is modulated according to panel (a). We show the STA (solid black) and the TRA stroke (red dashed). The horizontal gray dotted line is the \mossy{adiabatic energy change $\langle W^{AD} \rangle$}.}
\label{strokes}
\end{figure}

In Fig.~\ref{strokes} (a), we show an example of a ramp of the interaction strength for compression, which corresponds to the pulse that would be applied to achieve one of the adiabatic strokes of the Otto-cycle. Clearly, the functional form of these modulations depends heavily on the total duration of the stroke, $T_f$ and in the adiabatic limit we find $g_{AD}(t)=(\pi^2 a^4(t)-4)/2Na(t) \simeq -2/N a(t)$ (red, dashed) from Eq. (\ref{adiabticlimit}), which is a monotonic function. Conversely, for $T_f\!\!=\!\!0.15$ (solid, black) the ramp designed by the STA is more complex, most notably exhibiting a change in slope. This clearly shows that despite achieving essentially the same final state, employing a STA can imply that a drastically different trajectory is followed in order to compensate for the short timescale. In what follows, we will fix $T_f$ and examine the performance of a transformation facilitated by the STA. For comparison we use the same ramp given by the adiabatic limit, but performed in the shorter time $T_f$, and we refer to this as the time rescaled adiabatic (TRA) stroke. This allows for a fair comparison since as $T_f$ increases, the two modulations coincide.

\section{Performance of Shortcuts to Adiabaticity for Soliton Compression}
\label{resultsSTA}
We begin by examining the work done during and after a compression from $g_i\!=\!-0.1$ to $g_f\!=\!-0.2$. Fixing $T_f\!=\!0.15$, we show in Fig.~\ref{strokes} (b) the work for the correctly engineered STA (black) and the TRA (red dashed) stroke. The horizontal dotted line indicates the \mossy{adiabatic energy change $\langle W^{AD} \rangle$, which is the work done in the perfect adiabatic case}. For both approaches, we see that the average work obtained at $T_f$ is different from $\langle W^{AD} \rangle$, which implies that a certain amount of irreversible work was done during each stroke. However, using the STA leads to a significantly smaller degree of irreversibility.

\begin{figure}[t]
\includegraphics[width=\columnwidth]{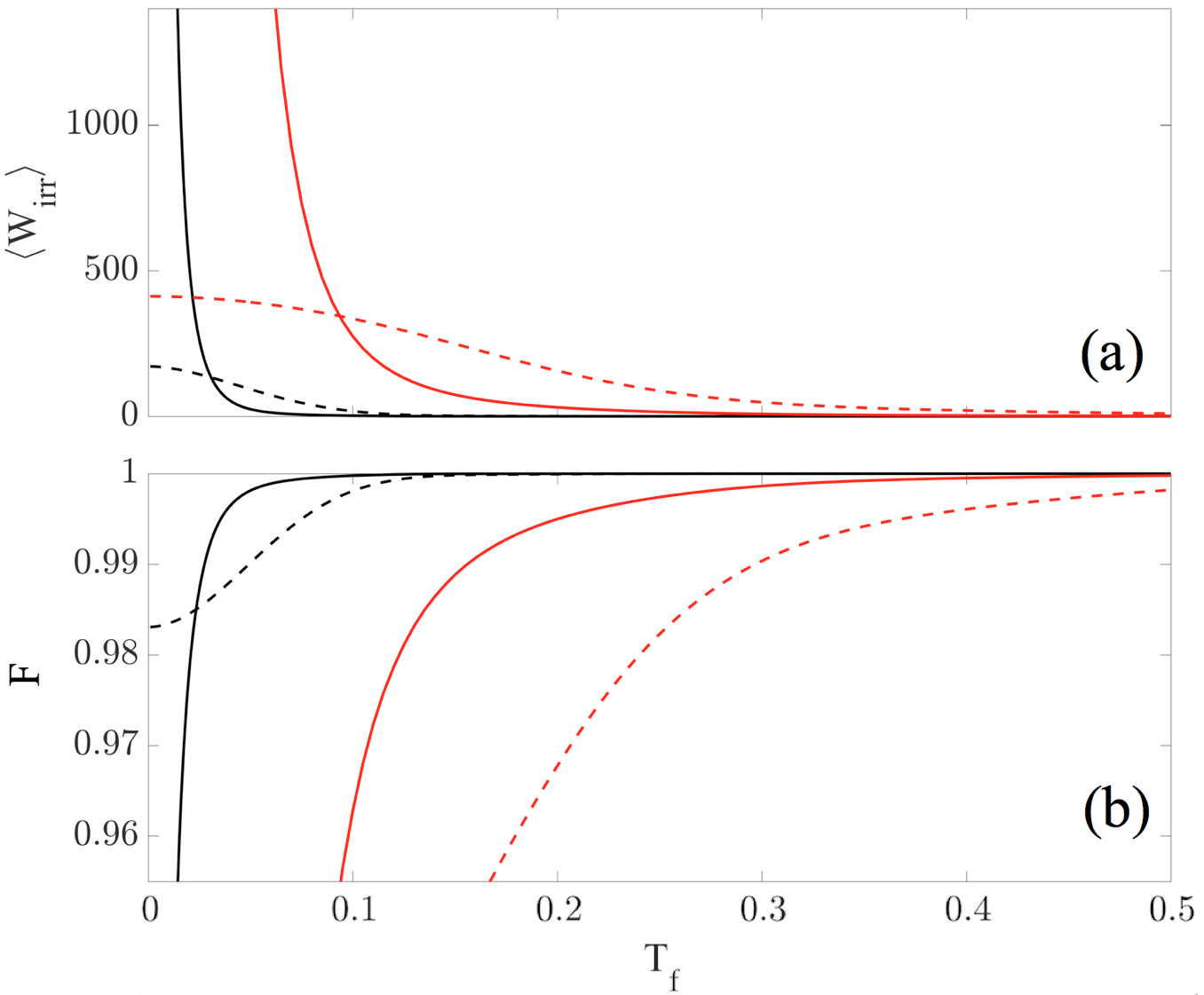}
\caption{(a) Average irreversible work, $\langle W_{irr}\rangle$ and (b) the final state fidelity, $F$ as a function of $T_f$. All panels show a compression of the soliton for $g_i\!=\!-0.1$ and $g_f\!=\!-0.2$ (red lines) and $g_i\!=\!-0.2$ and $g_f\!=\!-0.2646$ (black lines), with $N=100$. The solid lines represent the STA while the dashed lines are the TRA strokes.}
\label{fig:Fid_Wirr_Cost}
\end{figure}

In order to more clearly assess the relative performance of the two strokes, we show the irreversible work, as defined by Eq.~\eqref{eq:Wirr}, as a function of $T_f$ in Fig.~\ref{fig:Fid_Wirr_Cost} (a). Here the lines show $\langle W_{irr} \rangle$ at the end of the stroke for the STA (solid red) and TRA (dashed red) against $T_f$ for $g_i\!=\!-0.1$ and $g_f\!=\!-0.2$. Taking larger values of $T_f$ decreases $\langle W_{irr} \rangle$ for both stokes as we approach the adiabatic limit, whereas for small $T_f$ the amount of irreversibility created by both strokes increases. For $T_f \gtrsim 0.1$ the STA creates less irreversible work than the TRA as dynamical excitations are successfully suppressed, however for faster transformation times, $T_f\lesssim 0.1$, we find the converse. This can be understood by considering the modulations that are required by the STA for small $T_f$, cf. Fig.~\ref{strokes} (a). In this case the trajectory for $g(t)$ varies significantly, in stark contrast to the smooth, monotonically decreasing function used for the TRA stroke. This implies the need to input large amounts of energy during the stroke, in turn leading to a significant amount of irreversible work, which diverges as $T_f\to0$.

To investigate the role of the strength of the nonlinearity in these dynamics we show $\langle W_{irr} \rangle$ for stronger non-linear interaction strengths $g_i\!=\!-0.2$ and $g_f\!=\!-0.2646$ as the black curves in Fig.~\ref{fig:Fid_Wirr_Cost} (a). It is important to note that the energy difference between the initial and desired target states here is identical to the previous {\it weakly} non-linear case. However, we now see the remarkable role that non-linearities can play in engineering the STA. The black lines show that, while the qualitative behavior is consistent with the case of weakly non-linear interactions, the typical values of $\langle W_{irr} \rangle$ that can be achieved are lower. In fact, one can see that using stronger non-linear interactions allows for significantly faster strokes to be performed while still restricting the creation of excess excitations in the system. The reason for this is the effect that the increased non-linearity has on the energy spectrum. Larger non-linearity increases the gap between the energy eigenstates, which in turn allows for a faster driving since the system requires more energy to reach the excited states~\cite{CampbellPRA,CampbellPRL2}.

A crucial quantity in any control protocol is the final state fidelity, which allows us to quantify how close our final dynamical state at the end of the stroke, $\psi(x,T_f)$, is to the target equilibrium state, $\Psi(x)$,
\begin{equation}
F = |\langle \psi(x,T_f) ~ \Psi(x)\rangle |^2.
\end{equation}
In Fig.~\ref{fig:Fid_Wirr_Cost} (b) we show that the behavior of $F$ is consistent with the one of $\langle W_{irr} \rangle$, as the STA typically results in larger fidelities at the end of the stroke and therefore gives a more consistent approach to reach the target state than the TRA. Furthermore, using strong non-linear interactions allows for higher target fidelities at shorter stroke times, however in line with the irreversible work we can see that taking $T_f$ very small leads to the TRA stroke outperforming the STA. Nonetheless, in this case the actual fidelities are quite low in comparison to the desired target state rendering both approaches somewhat ineffective.

From Fig.~\ref{fig:Fid_Wirr_Cost} we learn three important points: {\it (i)} arbitrarily fast manipulation of the soliton matter-wave is not possible using this technique. Such fast manipulation leads to poor final fidelities with the target state, and comes accompanied by sizeable irreversible work. {\it (ii)} Stronger non-nonlinear interactions allow for faster strokes. By changing the gaps in the energy spectrum, the larger non-linear terms lead to better overall performance. {\it (iii)} For a realistic implementation, one could fix a minimum average post-stroke fidelity that state must achieve. This will then set a practical lower bound on $T_f$, which can then be used to determine optimal parameters in order to keep the irreversible work to a minimum.

\section{Performance Enhanced Otto Cycle with a Soliton Matter Wave}
\label{resultsENG}
While the above analysis only dealt with the compression of the matter-wave, qualitatively similar results hold for an expansion where the non-linear interaction strength is reduced. However, due to the effect the expansion has on the energy spectrum, this process generally results in slightly lower average fidelities and correspondingly higher irreversible work compared to soliton compression. Regardless, one can use the above approach to assess the performance of a quantum Otto-cycle facilitated using a shortcut to adiabaticity, as depicted in Fig.~\ref{schematic}. To this end, and in line with the analysis of Ref.~\cite{LutzArXiv,LutzArXiv2} where the case of the exactly solvable harmonic oscillator was treated, we calculate the efficiency of the cycle
\begin{equation}
\label{efficiency}
\eta=-\frac{\langle W_C \rangle+\langle W_E \rangle}{\langle Q_{N_-} \rangle}\;,
\end{equation}
where $\langle W_{C(E)} \rangle$ is the work during the compression (expansion) stroke. Here $\langle Q_{N_-} \rangle=\epsilon_E(0,N_E)-\tilde{\epsilon}_C(T_f, N_C)$ is the energy \mossy{change when particles are lost from the soliton to the free thermal gas}, where $\epsilon_E(0,N_E)$ is the initial energy of the soliton before the expansion stroke with $N_E$ particles, while $\tilde{\epsilon}_C(T_f, N_C)$ is the non-adiabatic energy at the end of the compression stroke with $N_C$ particles. \mossy{Therefore, more particles are condensed in the soliton for the compression than the expansion, $N_C>N_E$, so that when simulating the cycle the effect of temperature is captured by the difference between $N_C$ and $N_E$. As the normalization of the soliton wavefunction is dependent on $N_{C,E}$, the energy of the soliton will be modified by the change in particle number at end of the work strokes. This is needed to ensure power output from the engine cycle and gives the conditions that $\langle W_E \rangle + \langle W_C \rangle<0$ and $\langle Q_{N_-}\rangle>0$.} We can immediately see from Eq.~\eqref{efficiency} as we reduce the time to perform the strokes, $T_f$, the efficiency will decrease due to the growing irreversible work created during the STA,  cf.~Fig.~\ref{fig:Fid_Wirr_Cost} (a). We also assess the power generated
\begin{figure}[t]
\includegraphics[width=0.98\columnwidth]{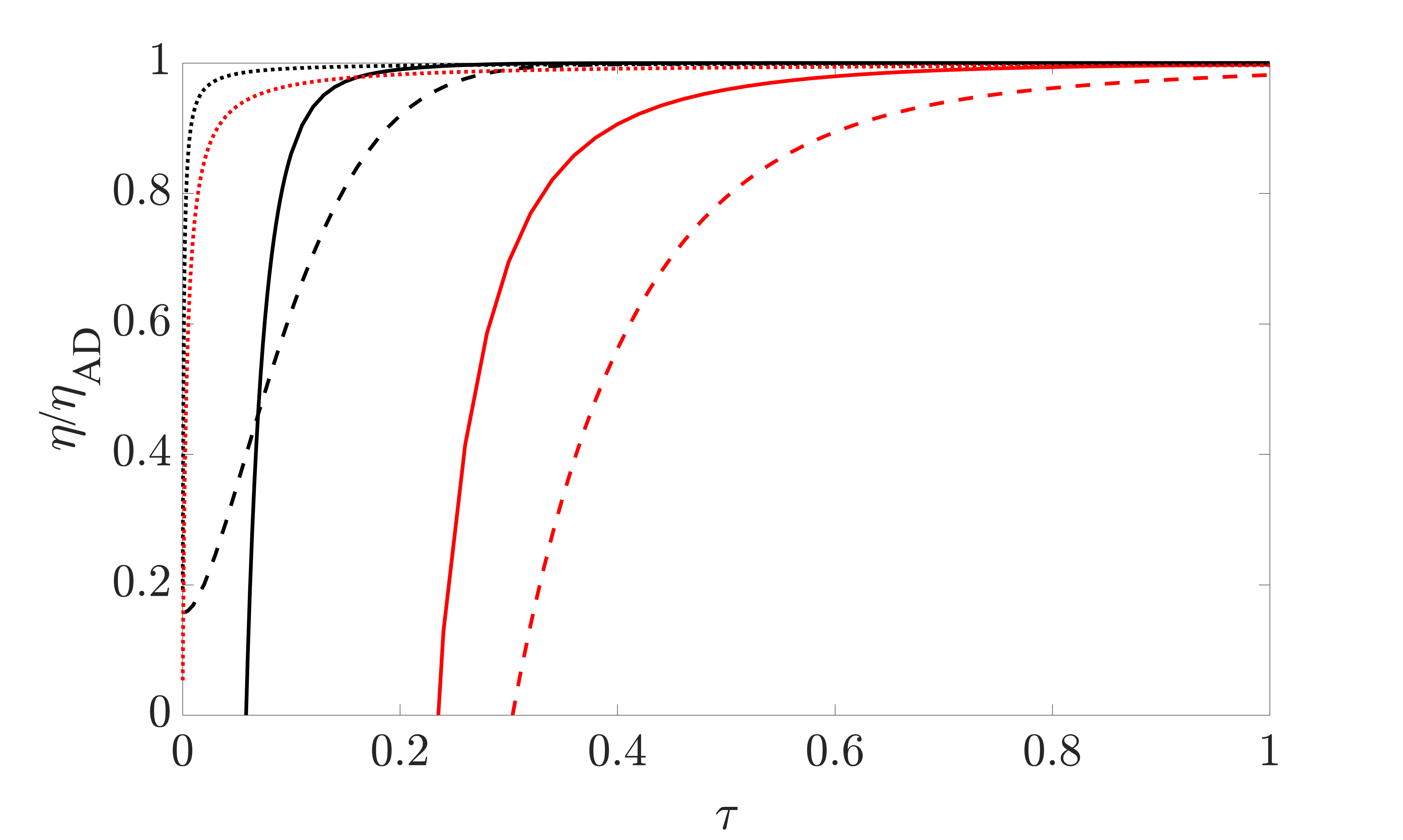}
\includegraphics[width=0.98\columnwidth]{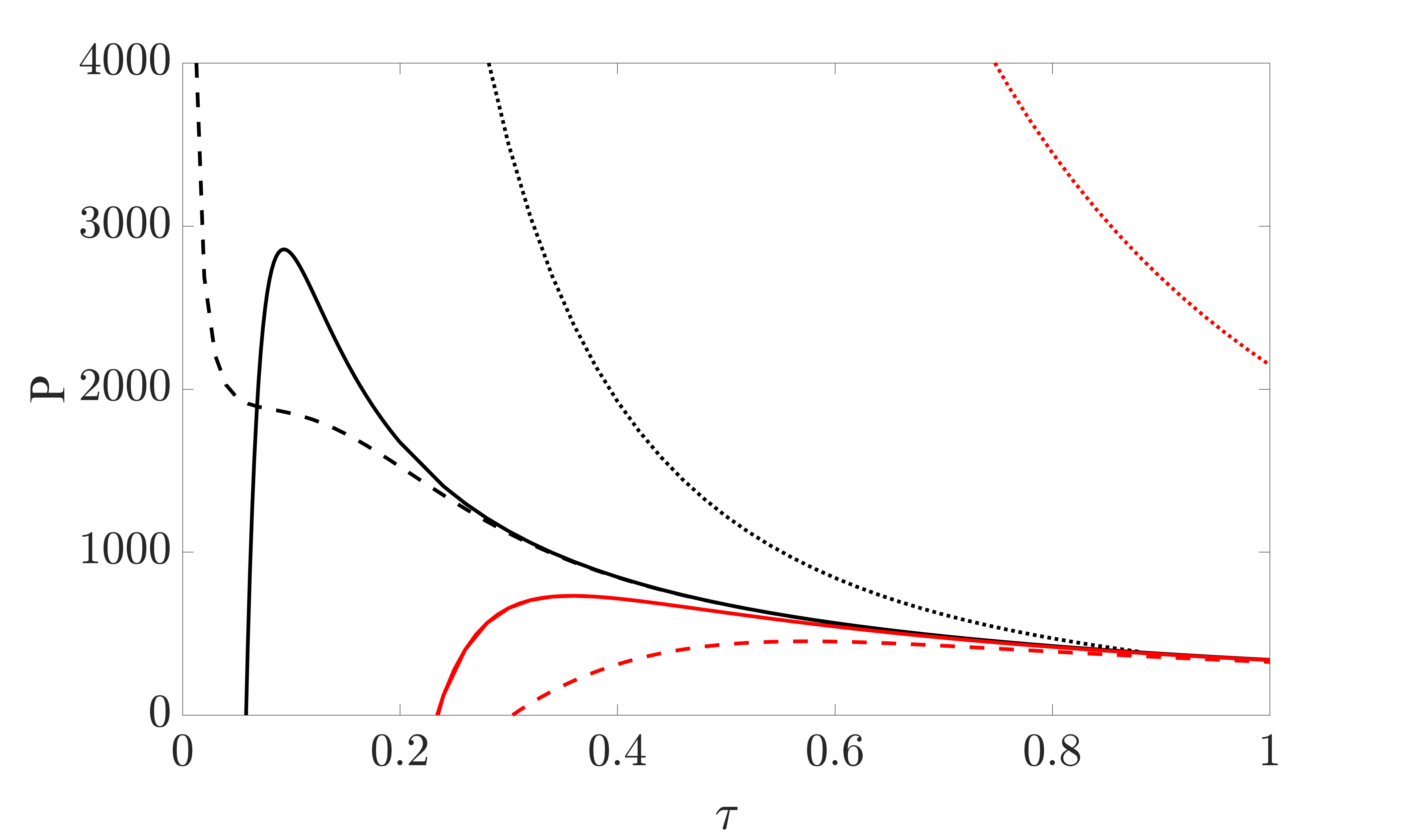}
\caption{(a) Efficiency of the cycle normalized with respect to the adiabatic efficiency $\eta_{\text{AD}}$ and (b) power, for the weakly non-linear case with $g_{i}\!=\!-0.1$ and $g_f\!=\!-0.2$ (red lines, $\eta_{\text{AD}}\approx0.76$) and strongly nonlinear case $g_{i}\!=\!-0.2$ and $g_{f}\!=\!-0.2646$ (black lines, $\eta_{\text{AD}}\approx0.43$). The solid lines represent the STA while the dashed lines are the corresponding TRA stroke. The thin dotted lines are the respective upper bounds dictated by the QSL. Along the compression stroke we take $N_C=100$ while along the expansion stroke we fix $N_E=90$.}
\label{fig:eff}
\end{figure}
\begin{equation}
P=-\frac{\langle W_C\rangle+\langle W_E \rangle}{\tau}\;,
\label{power}
\end{equation}
where $\tau$ is the total time for the whole cycle to be completed. We will assume that the time for the thermalization strokes, where \mossy{particles are} introduced or removed from the system, is much shorter than the time taken for the other strokes to be completed, and therefore $\tau\approx 2T_f$~\cite{LutzArXiv,LutzArXiv2,GooldSciRep}. 

Before analysing these quantities in detail, it is interesting to note that there exists upper bounds on the efficiency and the power by virtue of the quantum speed limit (QSL), which sets a lower bound on the time required for a quantum state to evolve~\cite{QSL_MT,QSL_ML,QSLDeffner,CampbellPRL2,FunoPRL} (see Ref.~\cite{QSLreview,DeffnerReview} for recent reviews). \mossy{We remark that, although we work with a mean-field dynamics, the QSL is known to extend to classical settings~\cite{DeffnerNJP, CSLarXiv1, CSLarXiv2}.} In the present context we can define the QSL as
\begin{equation}
T_f>T^{QSL}=\frac{\hbar B}{\langle \mathcal{E}_{STA} \rangle}\;,
\end{equation}
where $B$ is the Bures angle between the initial and final states~\cite{QSLDeffner} and $\langle \mathcal{E}_{STA} \rangle = \frac{1}{T_f} \int_{0}^{T_f} [\epsilon^I_{STA}(t) - \epsilon^I_{TRA}(t)] dt$ is the energy of the shortcut, where $\epsilon^I_{STA}(t)$ ($\epsilon^I_{TRA}(t)$) is the energy of the instantaneous eigenstate for the value of $g(t)$ for the STA (TRA). As shown in Ref.~\cite{LutzArXiv,LutzArXiv2} the upper bounds are then given by
\begin{equation}
\eta_{QSL}=-\frac{\langle W^{AD}_C \rangle+\langle W^{AD}_E \rangle}{\langle Q_{N_-} \rangle + \hbar(B_C+B_E)/\tau}
\end{equation}
\begin{equation}
P_{QSL}=-\frac{\langle W^{AD}_C \rangle+\langle W^{AD}_E \rangle}{T_C^{QSL}+T_E^{QSL}}
\end{equation}
where $B_C$ and $B_E$ are the the Bures angles for the compression and expansion strokes, and \mossy{$\langle W_{C,E}^{AD} \rangle$ is the adiabatic work difference for compression or expansion of the soliton}. The thin dotted curves in Fig.~\ref{fig:eff} correspond to these upper bounds.

We examine the efficiency and power of the Otto cycle in Fig.~\ref{fig:eff} for weak (red) and strong (black) non-linear interaction strengths for the STA (solid) and the TRA (dashed) strokes, \mossy{while the compression and expansion strokes are implemented with solitons composed of $N_C=100$ and $N_E=90$ particles respectively}. As done previously, to ensure a fair comparison, the magnitude of the change in the non-linear interaction is modified such that in the adiabatic limit the work performed during each stroke of the two realizations is the same. Furthermore, the power in the adiabatic limit for the two interaction regimes will also be identical and we rescale $\eta$ with respect to the adiabatic efficiency, $\eta_{AD}$. We see that the efficiency is always poor for small $T_f$, regardless of the interaction regime or the type of ramp applied. Larger $T_f$ allows one to realize a significantly more efficient cycle, one that operates at close to the adiabatic efficiency and, on these timescales, we find the use of the STA is always advantageous. Furthermore, we again see that stronger non-linearities lead to a better overall performance, while the output power is more sharply peaked and experiences a sharp cutoff. This is a consequence of the increased slope of $\langle W_{irr} \rangle$ at short $T_f$, as the designed STA must approach $g\rightarrow0$ to compensate for the excess energy put into the system. For $|gN|\approx 1$ the $sech$ ansatz for the soliton breaks down and the STA becomes ineffective. Finally, comparing with the bounds defined by the QSL, we see that for intermediate timescales strong non-linearities allow for an efficiency and power close to maximal. As we increase $T_f$, and all protocols approach the adiabatic limit, we find the curves all converge on top of one-another.

\begin{figure}
\includegraphics[width=0.98\columnwidth]{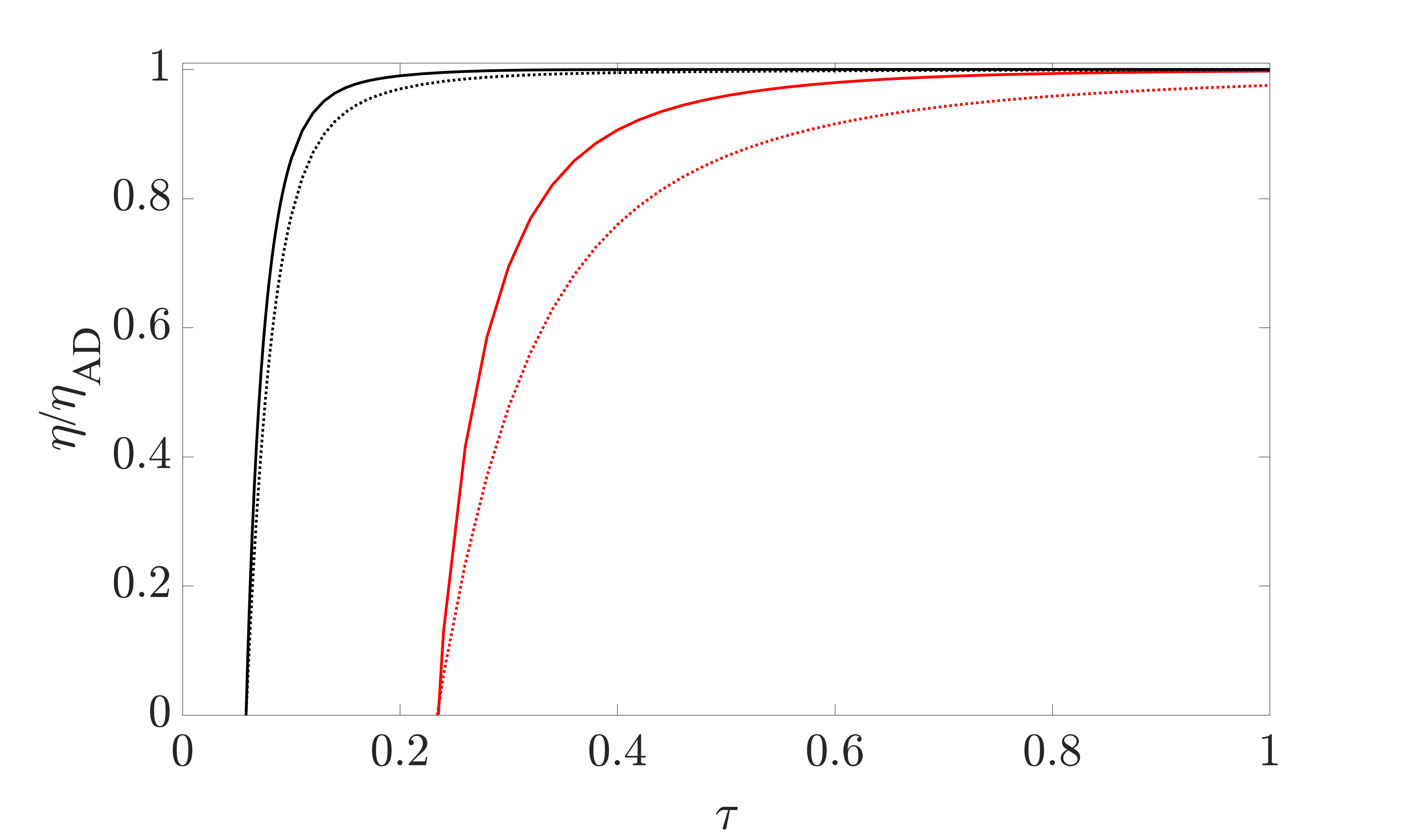}
\includegraphics[width=0.98\columnwidth]{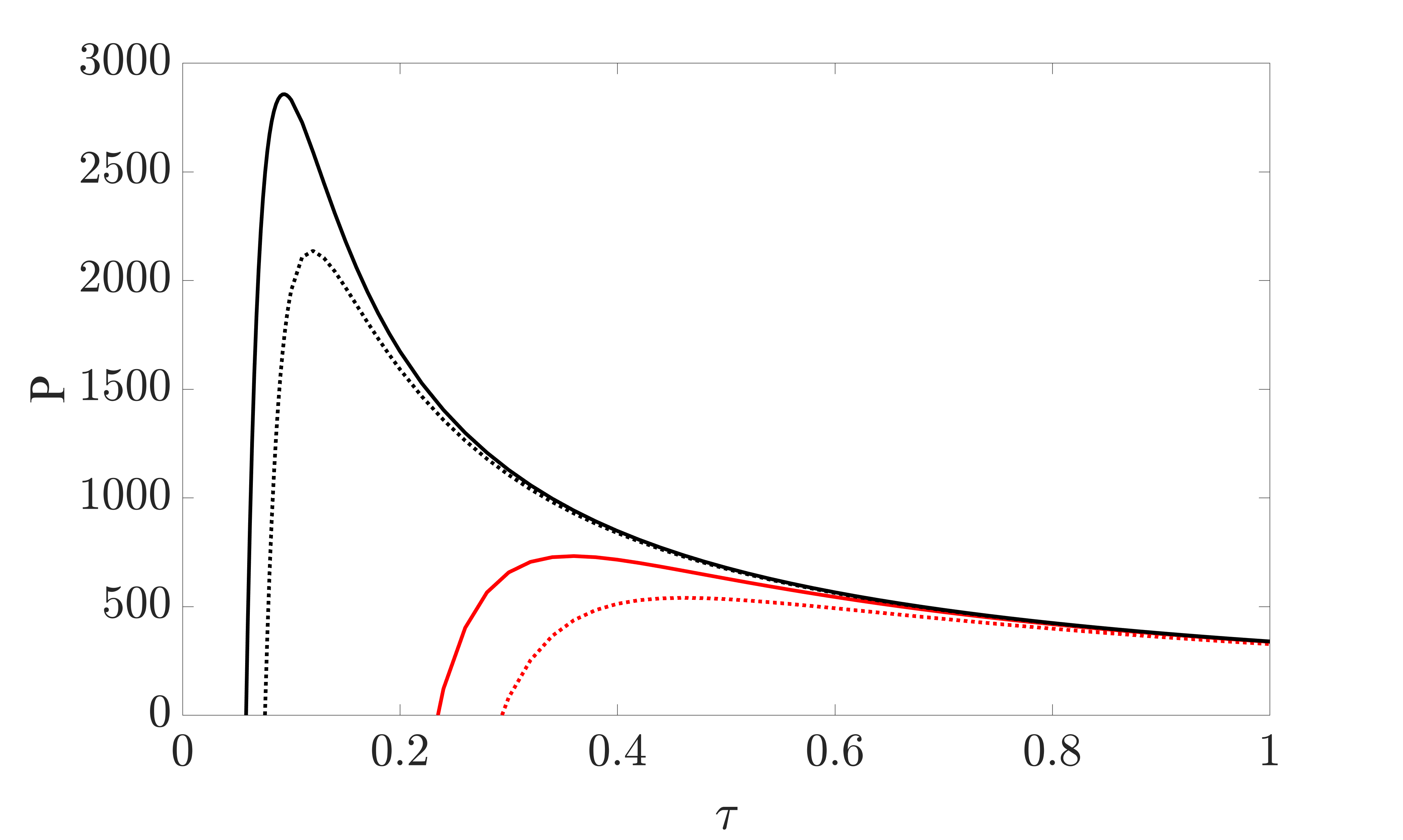}
\caption{(a) Efficiency of the cycle normalized with respect to the adiabatic efficiency $\eta_{\text{AD}}$ and (b) power, for the weakly non-linear case with $g_{i}\!=\!-0.1$ and $g_f\!=\!-0.2$ (red lines, $\eta_{\text{AD}}\approx0.76$) and strongly nonlinear case $g_{i}\!=\!-0.2$ and $g_{f}\!=\!-0.2646$ (black lines, $\eta_{\text{AD}}\approx0.43$). The solid lines represent efficiency and power calculated with Eq.~\eqref{efficiency} and Eq.~\eqref{power} respectively. The thin dotted lines are efficiency and power calculated with Eq.~\eqref{efficiencyC} and Eq.\eqref{powerC} respectively, which includes the added energy required to implement the STA.}
\label{fig:eff2}
\end{figure}

An important caveat must be stressed regarding the above results. Our definitions of efficiency and power do not account for any cost (energetic or otherwise) in achieving the desired dynamics. Such a question has received intense interest recently~\cite{CampbellPRA, Ecut, LutzArXiv, LutzArXiv2, CampbellPRL2, FunoPRL, Kosloff2017, SantosSciRep}. A reasonable (although not unique) definition for the cost of achieving one of the expansion/compression strokes is to determine the energy required for the pulse, i.e. $\big< \mathcal{E}_{STA} \big>$. As this represents an additional energy input, the efficiency then becomes
\begin{equation}
\eta_{cost}=-\frac{\langle W_C \rangle+\langle W_E \rangle}{\langle Q_{N_-} \rangle + \langle \mathcal{E}_{STA} \rangle_{C}+\langle \mathcal{E}_{STA} \rangle_{E}}.
\label{efficiencyC}
\end{equation}
Including this additional energy term should also have an effect on the output power. Indeed, if we view the use of the STA as a means to boost performance, the added power inputted through the use of the pulse should be subtracted from the total output power, thus
\begin{equation}
P_{cost}=-\frac{\langle W_C \rangle+\langle W_E \rangle-\langle \mathcal{E}_{STA} \rangle_{C}-\langle \mathcal{E}_{STA} \rangle_{E}}{\tau}\;.
\label{powerC}
\end{equation}
In Fig.~\ref{fig:eff2} we compare these quantities to Eqs.~\eqref{efficiency} and \eqref{power}. Clearly, the qualitative behavior is consistent, with the overall effect to slightly decrease the performance. However, we see that the advantages pointed out previously still persist, allowing us to conclude that even by including the additional energy required to achieve the dynamics, the STA can still significantly boost the performance of the cycle. As a final remark, while there are several (related) definitions for the cost required to achieve a STA currently in the literature~\cite{CampbellPRA, Ecut, LutzArXiv, LutzArXiv2, CampbellPRL2, FunoPRL, Kosloff2017, SantosSciRep}, regardless of which approach is chosen the general features outlined here will persist.

\section{Conclusions}
\label{conclusions}
We have analyzed the performance of a recently proposed shortcut to adiabaticity (STA) which modulates the non-linear interaction of a soliton matter wave. We quantified the effectiveness of the STA during compression of the soliton by calculating the irreversible work and the fidelity of the final state. We showed that the STA is a viable technique to efficiently suppresses excitations on non-adiabatic timescales, while its use on arbitrarily short timescales results in the generation of a significant degree of irreversibility when implementing the STA. Examining the performance of an Otto cycle using the soliton matter-wave as a working substance, we have shown that the STA can be a useful tool for these intermediate timescales, and that larger non-linear interaction strengths lead to a better overall performance. Our results thus significantly add to the study of quantum thermal cycles by taking advantage of the versatility of BECs to create a Feshbach engine, which can be efficiently controlled by tuning the non-linearity according to the STA. This system is also experimentally viable, where the energy of the soliton can be extracted from in-situ observations of the density, or through time-of-flight measurements of the momentum distribution.

\acknowledgements
We thank Sebastian Deffner for insightful discussions. This work was supported by the Okinawa Institute of Science and Technology Graduate University. We acknowledge support from the NSFC (11474193), the Shuguang program (14SG35), the Program for Professor of Special Appointment (Eastern Scholar), and the COST Action MP1209 ``Thermodynamics in the Quantum Regime".


\bibliography{feshbach_engine}

\end{document}